\documentclass[aps,prb,twocolumn,groupedaddress,showpacs]{revtex4}
\usepackage{graphicx}
\usepackage{dcolumn}
\usepackage{bm}
\usepackage{amsmath,amssymb}

\begin{document}



\title{
The second law of thermodynamics 
from concave energy in classical mechanics}  
\author{
C. Itoi 
and M. Amano}
\affiliation{
Department of Physics,  GS $\&$ CST,
Nihon University, \\Kandasurugadai, Chiyoda, 
  Tokyo 101-8308, Japan}

\date{\today}
\begin{abstract}
 A recently proposed quantum mechanical criterion `concavity of energy' for the second law of thermodynamics is studied also for 
classical particle systems confined in a bounded region by a potential with a time-dependent coupling constant.
 It is  shown that the time average 
 of work done by particles
 in a quench process cannot exceed that in the corresponding quasi-static process, if 
 the energy is a concave  function of the coupling constant. 
 It is proven that the energy is indeed concave for a general confining potential with  certain properties. 
This result implies that the system satisfies the  principle of maximum work in the adiabatic environment 
as an expression of the second law of thermodynamics.  
\end{abstract}
\maketitle

\section{Introduction}
 The impossibility to make any perpetual motion of second kind is 
 an expression of the second law of thermodynamics. 
This impossibility is expected to be clarified  within the framework of classical or quantum  mechanics. 
Planck's  principle  is well-known as a specific expression of this impossibility.    
This principle claims  that a positive amount of work cannot be extracted  from  
thermodynamic systems in an arbitrary cyclic operation in the adiabatic environment.     
 Any realization of a perpetual motion of second kind implies the violation of this principle. 
 Passivity is known as  a sufficient condition on a quantum state  for the impossibility of work extraction from quantum systems.  
We say that a quantum state is passive, if a positive amount of work cannot be extracted from this state as an initial state 
in any cyclic unitary evolution. 
Some mixed states, such as the Gibbs states are passive  \cite{PW,L}, but general pure states except the ground state are not passive.
The Gibbs distributions of classical  dynamical  systems  have a similar property to passive quantum states.
The well-known Jarzynski equality for classical dynamical systems enables us to show that
 the amount of work done by a classical dynamical system with an initial Gibbs distribution  
  in a general process  cannot exceed that in the 
quasi-static process \cite{J}.   This property of the Gibbs distribution corresponds  to the principle of maximum work in an isothermal environment
in thermodynamics. 
 
 On the other hand, there have been several purely mechanical studies on the work extraction by a physical evolution  from an initial  pure state  
  in an adiabatic environment within the frame work of quantum mechanics
  \cite{DCHFGV,GRE,PRGWE,MR,LLMPP,JSDMCG,SKCSG,DPZ,KIS}.   From the view point of the second law of thermodynamics,
 Kaneko, Ioda and Sagawa  study  Planck's principle from the view point of 
  the eigenstate thermalization hypothesis (ETH) analogue in quantum spin systems  \cite{KIS}.
 They give  numerical evaluations of  amounts of works from  energy eigenstates of quantum spin systems in a composite cyclic process 
 which consists of  instantaneous changes of the interaction 
 and leaving the system a while.  They find the strong ETH like behavior  only in non-integrable systems, where
 negative amounts of work are extracted  from every  energy eigenstate  in an cyclic process.
 In integrable systems, however,  they find only weak ETH behavior.    
 The principle of maximum work in adiabatic environment  
is  also well-known as an equivalent  principle to  Planck's  principle.
  This principle  states that work done by the 
 system in an arbitrary process cannot exceed the work in the corresponding quasi-static process. 
 A criterion `concave energy eigenvalues'  for quantum systems to satisfy this principle has been proposed, and checked 
 in quantum dynamical systems \cite{IA}.  It is proven that  an energy eigenstate expectation value of 
work done by the system in a quench process cannot exceed the work in the corresponding quasi-static process in adiabatic environment, 
if its energy eigenvalue is a concave function of the coupling constant. 
The concavity of energy eigenvalues is checked in several specific quantum systems.
This concavity criterion reduces the problem of  time-dependent quantum dynamics to a condition on the  time-independent quantum mechanics.  
 
  In the present paper,  we check a proposed  new criterion to satisfy 
 the principle of maximum work in the adiabatic environment also for classical particle systems.
 Since the second law is valid irrespective of woking substance in thermodynamics, 
 its validity  should be checked for the ideal gas as a simple example.   Despite  the well-known fact that interactions and 
 non-integrability of systems
 are essential for thermalization of mechanical systems,
 the impossibility of perpetual motion of second kind should be valid for free particles.
 We study a classical dynamics of non-interacting particles confined in bounded region
 of one dimensional space by a time-dependent confining potential. 
 We prove the concavity of energy as a function of the coupling constant is necessary 
  also for classical particle systems to satisfy the principle of maximum work.
 We evaluate difference of work between quench and quasi-static changes of the coupling.
  Since the quench change of the coupling constant is operated  instantaneously,  
  the lost energy of the particle in the quench process depends on the position of the particle when
  the coupling  is changed. On the other hand, the amount of work is unique  in the quasi-static process.  
 We employ the time average of the work over the oscillation period of the confined particle, and 
 prove that this average work in the quench process
  cannot exceed the work in the quasi-static process. 
  This agrees with the maximal work principle in the adiabatic environment as an expression of 
 the second law of  thermodynamics, which is equivalent to Planck's principle.

\section{Average of particle energy}
Consider  a particle with a unit  mass in one dimensional space, and let $x \in {\mathbb R}$ be its position coordinate.
This particle  is confined in a bounded region by a confining potential $\lambda u(x) \geq 0$ with  a coupling constant $\lambda \geq 0$.
Assume that 
(1) $u'(x)$ and $u''(x)$ exist at any $x \in {\mathbb R}$,
(2) $u'(x) \leq 0$ for $x <0$ and  $u'(x) \geq 0$ for $x>0$,
(3)
$u(-x) = u(x).$
 The equation of motion is given by
 \begin{equation}
 \ddot x = -\lambda u'(x).
 \end{equation}
The kinetic energy 
$$
K=\frac{1}{2} \dot{x}^2,
$$ 
and potential gives the energy of the particle
\begin{equation}
E=K + \lambda u(x),
\label{energy}
\end{equation} 
which is a constant of the motion if $\lambda$ is a constant.
One can imagine that the particle's motion becomes  a periodic oscillation around the origin $x=0$ for a constant $\lambda$.

Consider the case that $\lambda(t)$ is a 
function of a time parameter
 $t \in {\mathbb R}$.
In this case the energy $E_t$ is also a function of the time $t \in {\mathbb R}$.
In the present paper, we compare work done by the particle  in a quasi-static process and a quench process.
First, we consider a quasi-static process. 
Assume an initial condition $(x(0),   \dot x(0))=(\alpha_0, 0)$ 
with the coupling $\lambda(0)=\lambda_0$, and the initial energy is given by $E_0 = \lambda_0 u(\alpha_0)$. 
At an arbitrary time $t >0$,  the amplitude $\alpha_t$ determines  the energy $E_t=\lambda(t) u(\alpha_t)$.
The quasi-static process is given by an infinitesimally slow change of the coupling constant $\lambda(t)$.  
Assume that 
 the motion of the particle is a periodic oscillation around the origin $x=0$ with an amplitude $\alpha_t$.
For an arbitrary quasi-static process, the adiabatic invariant 
\begin{equation}
I =2 \int_{-\alpha_t} ^{\alpha_t} \sqrt{2(\lambda(t) u(\alpha_t)-\lambda(t) u(x))} dx, 
\label{adinv}
\end{equation}
is known as a constant of motion for  the slowly varying coupling constant $\lambda(t)$ \cite{LL}.
Here, we regard the amplitude  $\alpha(\lambda)$ as an  implicit
 function of $\lambda$, such that  the adiabatic invariant $I$  defined by
 \begin{equation}
I = 2\int_{-\alpha} ^{\alpha} \sqrt{2\lambda u(\alpha)-2\lambda u(x)} dx, 
\label{adinv2}
\end{equation}
 is a constant independent of $\lambda$ and define the energy 
 \begin{equation}E(\lambda) = \lambda u(\alpha(\lambda)),
 \label{energyfunction}
 \end{equation} 
 as a function of $\lambda$.  Note that 
 $$
 \alpha_t = \alpha(\lambda(t)), \ \ \ 
 E_t = E(\lambda(t)).
 $$
 Under the quasi-static change $\lambda_0\to\lambda_1$, the change of energy $\Delta E = E(\lambda_1) -E(\lambda_0) $ of the particle  
 gives the work $W=-\Delta E$ done by the particle in this process. 
 
Next, to consider a quench process,
define the period of the oscillating particle  
as a function of a coupling $\lambda$ by
\begin{equation}
T(\lambda)=2\int_{-\alpha} ^{\alpha} \frac{dx}{\dot x}=
 \int_{-\alpha} ^{\alpha} \frac{2 }{\sqrt{2(\lambda u(\alpha)-\lambda u(x))}} dx.
 \label{period}
\end{equation}
Define a function $\tau$, which is a spending time  of the particle in an interval $(a, b) \subset {\mathbb R}$  in the one dimensional space
during the period $T(\lambda)$ 
\begin{equation}
\tau(a,b,\lambda) := \int_a ^b \frac{2}{\sqrt{2(E(\lambda)-\lambda u(x))}} dx,
\end{equation}
as a function of $a, b$ and the coupling constant $\lambda$.  Note that the period is represented in terms of $\tau$
$$
T(\lambda) = \tau(-\alpha(\lambda), \alpha(\lambda),\lambda).
$$
Let us regard the position coordinate $x$ of the  particle  as a random variable,  and
define the probability for $x \in (a, b)$ by a ratio of particle's spending time in the interval $(a, b)$ 
\begin{equation}
P[x \in (a,b)] := \frac{\tau (a,b,\lambda)}{T(\lambda)}.
\end{equation}  
For a particle  with  an initial condition $(x(0), \dot x(0))=(\alpha, 0)$ in a given potential $\lambda u(x)$, 
above definition of the probability gives
the following  probability density function $p(x, \alpha)$ of $x$
\begin{equation}
p(x,\alpha) := 
 \frac{2}{T(\lambda)\sqrt{2(\lambda u(\alpha)-\lambda u(x))}}.
\end{equation} 
 Consider a quench process  given by an instantaneous change $\lambda_0 \to  \lambda_1$  of the coupling constant $\lambda$. 
 If this instantaneous change occurs at time $t$, the change of the particle  energy is given by $(\lambda_1-\lambda_0) u (x(t))$. 
 The work $-(\lambda_1-\lambda_0) u (x(t))$ done by particle in this quench process depends on $t$.   
To evaluate the work done by the particle in the quench process, we employ the average of $u(x(t))$  
over the oscillation period $T(\lambda_0)$ represented in terms of the integration over $x \in (-\alpha_0, \alpha_0)$
\begin{eqnarray}
&&\bar{u} (\lambda_0)=\frac{2}{T(\lambda_0)} \int_0^{\frac{T(\lambda_0)}{2}} u(x(t))dt\\
&&= \int_{-\alpha_0}^{\alpha_0} u(x)p(x,\alpha_0)dx \\
&&=\frac{2}{T(\lambda_0)} \int_{-\alpha_0} ^{\alpha_0} \frac{ u(x)}{\sqrt{2(E(\lambda_0)-\lambda_0 u(x))}} dx,
\label{av}
\end{eqnarray}
Note that the average $\bar u (\lambda)$ in this probability
 is given by the derivative $E'(\lambda)$ of the function $E(\lambda)$ 
defined by (\ref{adinv2})
and (\ref{energyfunction})  implicitly  
\begin{equation}
E'(\lambda)=\bar  u (\lambda).
\label{E'}
\end{equation}   
This formula is a classical analogue to the Hellmann-Feynman theorem in quantum mechanics \cite{H,F}.
Therefore, the average work done by the quench process $\lambda_0\to\lambda_1$ is 
\begin{equation}
(\lambda_0-\lambda_1) E'(\lambda_0).
\end{equation}
If the quasi-static process gives the maximal work,  we have the following inequality
 \begin{equation}
(\lambda_0-\lambda_1) E'(\lambda_0) \leq E(\lambda_0) -E(\lambda_1).
\label{ineq}
\end{equation}
This implies the concavity of the function $E(\lambda)$.   
 
Next, we consider $N$ independent particles confined by the same confining potential.
The mechanical state is described by a set   $x=(x_1, x_2, \cdots, x_N) \in {\mathbb R}^N$ of positions. 
The total energy ${\cal E}$ is given by a summation of energy  over all particles with a
set $\alpha(\lambda):=(\alpha_1(\lambda), \alpha_2(\lambda), \cdots, \alpha_N(\lambda)) \in {\mathbb R}^N$ of  amplitudes 
\begin{equation}
{\cal E}(\alpha(\lambda)) := \sum_{n=1} ^N E_n(\lambda), 
\end{equation}
where the energy of $n$-th particle is defined by $E_n(\lambda):= \lambda u(\alpha_n(\lambda)).$
Since these particles are independent, the probability density function of $x \in {\mathbb R}^N$ is given by
a product of that of the position coordinate of each particle  
$$
P(x,\alpha(\lambda)) := \prod_{n=1}^N p(x_n,\alpha_n(\lambda)).
$$ 
Then, the average work done by $N$ particles in a quench process $\lambda_0\to\lambda_1$ is 
\begin{equation}
(\lambda_0-\lambda_1) \sum_{n=1} ^N \bar u(\alpha_n(\lambda_0))=(\lambda_0-\lambda_1) \sum_{n=1} ^N E_n'(\lambda).
\end{equation}
 The work in the corresponding quasi-static process is 
 $${\cal E}(\alpha(\lambda_0))-{\cal E}(\alpha(\lambda_1)) = \sum_{n=1}^N (E_n(\lambda_0)-E_n(\lambda_1)).
 $$  
Therefore,  the work in quench process cannot exceed that in the corresponding quasi-static process, if 
$$
(\lambda_0-\lambda_1) E_n'(\lambda)\leq E_n(\lambda_0)-E_n(\lambda_1),
$$
which is guaranteed by the concavity of each $E_n(\lambda)$.
We prove the concavity of  a single particle energy in the following.

\section{Proof of concavity}
To prove the concavity of the function $E(\lambda)$,
 first we show that  increasing in $\lambda$  of  $\tau(0,\alpha, \lambda)$ implies
 the concavity of the energy function $E(\lambda)$ with $E(\lambda) = \lambda u(\alpha)$.
 The derivative of the energy function is 
 \begin{equation}
 E'(\lambda) = \bar u(\lambda) = \frac{E(\lambda) -\bar K(\lambda)}{\lambda}.
 \label{1stE}
 \end{equation}
 where the average of the kinetic energy as a function of $\lambda$ is defined  by
\begin{equation}
\bar K(\lambda) =\frac{\sqrt{2}}{T(\lambda)} 
\int_{-\alpha(\lambda)} ^{\alpha(\lambda) }\sqrt{E(\lambda)-\lambda u(x)} dx=\frac{I}{2T(\lambda)} .
\label{avK}
\end{equation} 
Note that the identities (\ref{1stE}) and (\ref{avK}) give
 the  second derivative $E''(\lambda)$ of the energy function
\begin{equation}
E''(\lambda) = -\frac{\bar K'(\lambda)}{\lambda} = \frac{I T'(\lambda)}{2\lambda T(\lambda)^2}.
\label{E2nd}
\end{equation}
This indicates that  the positive semi-definiteness of $\bar K'(\lambda)$ or  $-T'(\lambda)$
implies  the concavity of the energy function $E(\lambda)$.   
Intuitively,  it is obvious that $\bar K(\lambda)$ is monotonically increasing,  if $E(\lambda)$ is monotonically increasing. 
Since the change $\Delta E=E(\lambda_1)-E(\lambda_0)$  by a quasi-static operation
$\lambda_0\to\lambda_1$ is shared by $\Delta \bar K=\bar K(\lambda_1)-\bar K(\lambda_0)$ 
and $\Delta \bar U=\lambda_1 \bar u(\lambda_1) - \lambda_0 \bar u(\lambda_0)$
$$
\Delta E=\Delta \bar K+\Delta \bar U,
$$
then,  all signs of  $\Delta E$, $\Delta \bar K$ and $\Delta \bar U$
should be the same. 

 Let us prove the concavity of the energy function $E(\lambda)$.
To this end, we prove the non-positivity of  derivative $T'(\lambda)$ on $\lambda$.
This derivative has an apparent  divergence at $x=\alpha$ 
 in the following naive calculation
\begin{eqnarray}
&&\frac{1}{2} T'(\lambda)= \frac{d}{d\lambda} \tau(0,\alpha(\lambda),\lambda) \\
&&=\frac{\sqrt{2}\alpha'(\lambda)}{\sqrt{E(\lambda) -\lambda u(\alpha)}}-
\frac{1}{ \sqrt{2}}  \int_0 ^{\alpha(\lambda)} \frac{E'(\lambda)-u(x)}{(E(\lambda)-\lambda u(x))^{\frac{3}{2}} }dx, \nonumber
\end{eqnarray} 
since $E(\lambda)=\lambda u(\alpha(\lambda))$.
To regularize the integration at the  boundary  
$x=\alpha(\lambda)$, we decompose $\tau(0,\alpha(\lambda), \lambda)$ into the following three terms
\begin{eqnarray}
&&\tau(0,\alpha(\lambda),\lambda) \\
&&=\tau(0,c,\lambda) +
\tau(c, \alpha(\lambda)-\epsilon,\lambda)+ 
\tau(\alpha(\lambda)-\epsilon,\alpha(\lambda),\lambda),  \nonumber 
\label{decomposes}
\end{eqnarray}
where  $0  < c< \alpha(\lambda)-\epsilon < \alpha(\lambda) $,  and $u(c) < \bar u(\lambda)$ with $E(\lambda) = \lambda u(\alpha(\lambda))$.
We show that each term is monotonically decreasing.
Note  the following relation
\begin{equation}
E'(\lambda) = u(\alpha(\lambda)) + \lambda u'(\alpha(\lambda)) \alpha'(\lambda).
\end{equation}
The the derivative of the first term in  (\ref{decomposes}) is obviously non-positive 
\begin{equation}
\frac{d}{d\lambda} \tau(0,c,\lambda) =-\frac{1}{ \sqrt{2}}  \int_0 ^c \frac{\bar u (\lambda)-u(x)}{(E(\lambda)-\lambda u(x))^{\frac{3}{2}} }dx.
\end{equation} 
The derivative of the second term in  (\ref{decomposes}) is
\begin{eqnarray}
&&\frac{d}{d\lambda}\tau(c, \alpha(\lambda)-\epsilon, \lambda) \\
&&=\frac{\sqrt{2}\alpha'(\lambda)}{\sqrt{E(\lambda)-\lambda u(\alpha(\lambda)-\epsilon)}} \\
&&-\frac{1}{ \sqrt{2}}  \int_c ^{\alpha(\lambda)-\epsilon} \hspace{-4mm}\frac{E'(\lambda)-u(x)}{(E(\lambda)-\lambda u(x))^{\frac{3}{2}} }dx. 
\label{2nd}
\end{eqnarray}
 An integration by parts  in the second term  gives 
 \begin{eqnarray}
 &&\frac{1}{\sqrt{2}}\int_c ^{\alpha(\lambda)-\epsilon} \frac{E'(\lambda)-u(x)}{(E(\lambda)-\lambda u(x))^{\frac{3}{2}} }dx\\
 &&=\frac{1}{\sqrt{2}}\int_c ^{\alpha(\lambda)-\epsilon} \frac{E'(\lambda)-u(x)}{(E(\lambda)-\lambda u(x))^{\frac{3}{2}} }\frac{\lambda u'(x)}{\lambda u'(x)}dx\\
 && =\hspace{-2mm}\int_c ^{\alpha(\lambda)-\epsilon}\hspace{-1mm}\Big[ \frac{\sqrt{2}}{(E(\lambda)-\lambda u(x))^{\frac{1}{2}} }\Big]'\frac{E'(\lambda)-u(x)}{\lambda u'(x)}dx\\
 &&= \left[ \frac{\sqrt{2}}{(E(\lambda)-\lambda u(x))^{\frac{1}{2}} }\frac{E'(\lambda)-u(x)}{\lambda u'(x)} \right]_c^{\alpha(\lambda)-\epsilon} \label{2nddecompoess}\\ %
&&- \int_c ^{\alpha(\lambda)-\epsilon}\hspace{-5mm}\frac{\sqrt{2}}{(E(\lambda)-\lambda u(x))^{\frac{1}{2}} } \frac{d}{dx}  \frac{E'(\lambda)-u(x)}{\lambda u'(x)}dx.
\label{2nddecompoess2}
 \end{eqnarray}
There exists a bounded positive-valued  function  $f_1(\epsilon) \leq C_1 \sqrt{\epsilon}$ with a positive constant $C_1$ independent of $\epsilon$ 
, such that  (\ref{2nddecompoess}) is
\begin{eqnarray}
&&\left[ \frac{\sqrt{2}}{(E(\lambda)-\lambda u(x))^{\frac{1}{2}} }\frac{E'(\lambda)-u(x)}{\lambda u'(x)} \right]_c^{\alpha(\lambda)-\epsilon} \\
&&= \frac{\sqrt{2}[E'(\lambda)
-u(\alpha(\lambda)-\epsilon)]}{(E(\lambda)-\lambda u(\alpha(\lambda)-\epsilon))^{\frac{1}{2}} \lambda u'(\alpha(\lambda)-\epsilon)} 
\\&&-\frac{\sqrt{2}}{(E(\lambda)-\lambda u(c))^{\frac{1}{2}} }\frac{E'(\lambda)-u(c)}{\lambda u'(c)}\\
&&=\frac{\sqrt{2}\alpha'(\lambda)}{\sqrt{E(\lambda)-\lambda u(\alpha(\lambda)-\epsilon)}} + f_1(\epsilon)\\
&&-\frac{\sqrt{2}}{(E(\lambda)-\lambda u(c))^{\frac{1}{2}} }\frac{E'(\lambda)-u(c)}{\lambda u'(c)}.
\label{c,alpha-epsilon}
\end{eqnarray}
Therefore this and (\ref{2nd}) give 
\begin{eqnarray}
&&\frac{d}{d\lambda} \tau(c, \alpha(\lambda)-\epsilon, \lambda) \nonumber \\
&&=-f_1(\epsilon)+\frac{\sqrt{2}}{(E(\lambda)-\lambda u(c))^{\frac{1}{2}} }\frac{E'(\lambda)-u(c)}{\lambda u'(c)} 
\nonumber \\
&&+\int_c ^{\alpha(\lambda)-\epsilon}\hspace{-5mm}\frac{\sqrt{2}}{(E(\lambda)-\lambda u(x))^{\frac{1}{2}} } \frac{d}{dx}  \frac{E'(\lambda)-u(x)}{\lambda u'(x)}dx. 
\nonumber 
\end{eqnarray}
The mean value theorem for definite integrals implies that
there exists a positive number $c' \in (c, \alpha(\lambda)-\epsilon)$, such that  the final term in the above  is  
\begin{eqnarray}
&&\int_c ^{\alpha(\lambda)-\epsilon}\hspace{-5mm}\frac{\sqrt{2}}{(E(\lambda)-\lambda u(x))^{\frac{1}{2}} } \frac{d}{dx}  \frac{E'(\lambda)-u(x)}{\lambda u'(x)}dx \\
&&= \left[ \frac{E'(\lambda)-u(\alpha(\lambda)-\epsilon)}{\lambda u'(\alpha(\lambda)-\epsilon)} -  
\frac{E'(\lambda)-u(c)}{\lambda u'(c)}  \right]  \\
&&\times\frac{\sqrt{2}}{(E(\lambda)-\lambda u(c'))^{\frac{1}{2}} }.\nonumber
\end{eqnarray}
This gives  another expression  
\begin{eqnarray}
&&\frac{d}{d\lambda} \tau(c, \alpha(\lambda)-\epsilon, \lambda) \nonumber \\
&&=-f_1(\epsilon)+\frac{\sqrt{2}}{(E(\lambda)-\lambda u(c))^{\frac{1}{2}} }\frac{E'(\lambda)-u(c)}{\lambda u'(c)} 
\nonumber \\
&&+ \left[ \frac{E'(\lambda)-u(\alpha(\lambda)-\epsilon)}{\lambda u'(\alpha(\lambda)-\epsilon)}  
-\frac{E'(\lambda)-u(c)}{\lambda u'(c)}  \right] \nonumber \\
&&\times \frac{\sqrt{2}}{(E(\lambda)-\lambda u(c'))^{\frac{1}{2}} }\nonumber \\
&&= - f_1(\epsilon) -\frac{u(\alpha(\lambda)-\epsilon)-\bar u(\lambda)}{\lambda u'(\alpha(\lambda)-\epsilon)}\frac{\sqrt{2}}{(E(\lambda)-\lambda u(c'))^{\frac{1}{2}} }\label{2ndfinal}
\\&&-\hspace{-1mm} \left[\frac{\sqrt{2}}{(E(\lambda)-\lambda u(c'))^{\frac{1}{2}} } - \frac{\sqrt{2}}{(E(\lambda)-\lambda u(c))^{\frac{1}{2}} }\right] \frac{\bar u(\lambda)-u(c)}{\lambda u'(c)}. \nonumber
\end{eqnarray}
This is  negative semi-definite, since $u(c) \leq u(c')$.
Finally, we show that the  third term in (\ref{decomposes})  is infinitesimal.
An integration by parts in the  third term in (\ref{decomposes})  gives
\begin{eqnarray}
&&\tau(\alpha(\lambda)-\epsilon,\alpha(\lambda),\lambda) = 
\int_{\alpha(\lambda)-\epsilon} ^{\alpha(\lambda)} \frac{\sqrt{2}}{\sqrt{E(\lambda)-\lambda u(x)}} dx \nonumber \\
&&= \int_{\alpha(\lambda)-\epsilon} ^{\alpha(\lambda)} 
\frac{\sqrt{2}}{\sqrt{E(\lambda)-\lambda u(x)}} \frac{-\lambda u'(x)}{-\lambda u'(x)}dx \\
&&=\int_{\alpha(\lambda)-\epsilon} ^{\alpha(\lambda)} \frac{d}{dx}\Big[2 \sqrt{2(E(\lambda)-\lambda u(x))} \Big]\frac{-1}{\lambda u'(x)}dx \\
&&=\left[ \frac{2\sqrt{2(E(\lambda)-\lambda u(x))}}{-\lambda u'(x)}  \right]_{\alpha(\lambda)-\epsilon}^{\alpha(\lambda) }\\
&&- \int_{\alpha(\lambda)-\epsilon}^{\alpha(\lambda)} 2\sqrt{2(E(\lambda)-\lambda u(x)}) 
\frac{d}{dx} \frac{-1}{\lambda u'(x)} dx\\
&&= \frac{2\sqrt{2(E(\lambda)-\lambda u(\alpha(\lambda)-\epsilon))}}{\lambda u'(\alpha(\lambda)-\epsilon)} 
\\&&- \int_{\alpha(\lambda)-\epsilon}^{\alpha(\lambda)} 2\sqrt{2(E(\lambda)-\lambda u(x)}) 
\frac{d}{dx} \frac{-1}{\lambda u'(x)} dx.
\end{eqnarray}
The derivative of the above in $\lambda$ is
\begin{eqnarray}
&&\frac{d}{d\lambda} \tau (\alpha(\lambda)-\epsilon,\alpha(\lambda),\lambda) \\
&&=  \sqrt{2(E(\lambda)-\lambda u(\alpha(\lambda)-\epsilon))} \frac{\partial}{\partial \lambda}\frac{2}{\lambda u'(\alpha(\lambda)-\epsilon)}
\\&&+ \frac{\sqrt{2}(E'(\lambda) - u(\alpha(\lambda) -\epsilon) -\lambda u'(\alpha(\lambda)-\epsilon) \alpha'(\lambda))}{\lambda u'(\alpha(\lambda)-\epsilon) \sqrt{E(\lambda)-\lambda u(\alpha(\lambda)-\epsilon)}}\\
&&+\int_{\alpha(\lambda)-\epsilon} ^{\alpha(\lambda)}\left[\frac{2\sqrt{2(E(\lambda) -\lambda u(x))}}{\lambda^2} 
-\frac{2E'(\lambda)-2u(x)}{\lambda \sqrt{2(E(\lambda) - \lambda u(x))}} \right]
\nonumber \\ && \times
\frac{d}{dx}\frac{-1}{u'(x)} dx \\
&&+ \frac{2 \alpha'(\lambda) \sqrt{2(E(\lambda)-\lambda u(\alpha-\epsilon))} }{\lambda} \Big(\frac{1}{u'(\alpha(\lambda)-\epsilon)}\Big)' 
\\
&&=:f_2(\epsilon).
\end{eqnarray}
Taylor's theorem for a differentiable function $f$
implies that there exists $\theta \in (0,1)$, such that
$$
f(\alpha(\lambda)-\epsilon) = f(\alpha(\lambda))-\epsilon f'(\alpha(\lambda)-\theta \epsilon).
$$
Therefore,
there exists a positive number $C_2$ independent of $\epsilon$ such that  the absolute value of the  above  is bounded by 
\begin{equation}
|\frac{d}{d\lambda}\tau(\alpha(\lambda)-\epsilon,\alpha(\lambda),\lambda)| =
|f_2(\epsilon)| \leq C_2 \sqrt{\epsilon}.
\label{bounds3}
\end{equation}
Finally, we have
\begin{eqnarray}
&&\frac{d}{d\lambda}\tau(0, \alpha(\lambda), \lambda) \\
&&=\frac{d}{d\lambda}\tau(0,c,\lambda)+\frac{d}{d\lambda}\tau(c,\alpha(\lambda)-\epsilon,\lambda) \nonumber \\
&&+\frac{d}{d\lambda}\tau(\alpha(\lambda)-\epsilon,\alpha(\lambda),\lambda) \\
&&=-\frac{1}{ \sqrt{2}}  \int_0 ^c \frac{\bar u(\lambda)-u(x)}{(E(\lambda)-\lambda u(x))^{\frac{3}{2}} }dx \nonumber 
\\&&+\frac{\sqrt{2}}{(E(\lambda)-\lambda u(c))^{\frac{1}{2}} }\frac{E'(\lambda)-u(c)}{\lambda u'(c)} 
\nonumber \\&&+\int_c ^{\alpha(\lambda)-\epsilon}\hspace{-5mm}\frac{\sqrt{2}}{(E(\lambda)-\lambda u(x))^{\frac{1}{2}} } 
\frac{d}{dx}  \frac{E'(\lambda)-u(x)}{\lambda u'(x)}dx \nonumber\\&&-f_1(\epsilon) +f_2(\epsilon)\\
&&\leq  C_2\sqrt{\epsilon}-\frac{1}{ \sqrt{2}}  \int_0 ^c \frac{\bar u(\lambda)-u(x)}{(E(\lambda)-\lambda u(x))^{\frac{3}{2}} }dx \nonumber \\
&& -\left[\frac{\sqrt{2}}{(E(\lambda)-\lambda u(c'))^{\frac{1}{2}} } 
- \frac{\sqrt{2}}{(E(\lambda)-\lambda u(c))^{\frac{1}{2}} }\right]  \\
&&\times \frac{\bar u(\lambda)-u(c)}{\lambda u'(c)} \nonumber \\ 
&&-\frac{u(\alpha(\lambda)-\epsilon)
-\bar u(\lambda)}{\lambda u'(\alpha(\lambda)-\epsilon)} \frac{\sqrt{2}}{(E(\lambda)-\lambda u(c'))^{\frac{1}{2}}}.
\label{2ndfinal}
\end{eqnarray}
Since $u(c) \leq \bar u (\lambda) \leq u(c')\leq u(\alpha) $  and $\epsilon >0$ is arbitrary,
the above estimates give 
\begin{equation}
\frac{d}{d\lambda}\tau(0,\alpha(\lambda),\lambda) \leq 0.
\end{equation}
This and the relation (\ref{E2nd}) imply that the energy function $E(\lambda)$ is concave.
Therefore the work in the quench process $\lambda_0 \to \lambda_1$ cannot exceed 
that in the quasi-static process as given by the inequality (\ref{ineq}).

\section{Discussions}
First,   we summarize our results. We discuss classical dynamics of non-interacting particles confined by a time-dependent confining potential.
The energy function can be regarded as a function of the coupling constant, such that the adiabatic invariant is constant of motion 
with infinitesimally slow change of the coupling constant. We have shown that  
the work done by the particles in a quench process cannot exceed that in quasi-static process, if the energy is a concave function of the coupling constant.
We have proven the concavity  in a classical  single particle in a general potential, then also the energy of non-interacting particles is
concave.

Next, we discuss a sequential process and its convergence to the quasi-static process in classical dynamics.  Consider a composite process which consists of 
quench and waiting operations. Let $(\lambda_l)_{l=0,1,2, \cdots, L}$ be a sequence of coupling constants,  and consider $l$-th 
process which consists of quench operation $\lambda_{l-1} \to \lambda_{l}$ and leaving the system during a time interval $(t_{l-1}, t_{l})$. 
If the waiting time $t_{l-1}-t_{l-2}$ in the $(l-1)$-th process  is sufficiently long, the work done by the particle in the quench process $\lambda_{l-1}\to\lambda_{l}$
is given by $(\lambda_{l-1}-\lambda_{l})\bar u(\lambda_{l-1})=(\lambda_{l-1}-\lambda_{l})E'(\lambda_{l-1})$. 
Then, the total amount of the work done by the particle in this sequential process is
\begin{equation}
\sum_{l=1} ^{L}(\lambda_{l-1}-\lambda_{l}) E'(\lambda_{l-1}).
\end{equation} 
Let us consider the limit $L\to\infty$.
If $\lim_{L\to\infty} \lambda_L =\lambda_\infty $ converges and  the interval $\sup_l | \lambda_{l-1}-\lambda_{l}|\to 0$, 
the total amount of work converges to
\begin{eqnarray}
&&\lim_{L\to \infty} \sum_{l=1} ^{L}(\lambda_{l-1}-\lambda_{l})E'(\lambda_{l-1})\\&&
= -\int_{\lambda_0} ^{\lambda_\infty} E'(\lambda) d\lambda 
= E(\lambda_0) -E(\lambda_\infty),
\end{eqnarray} 
which is identical to the work in the quasi-static process $\lambda_0\to\lambda_\infty$.  Note that $t_L-t_0$ diverges as $L\to\infty$ to obtain the quasi-static process as this limit.
We consider that a general process with a finite operation speed can be constructed
by a certain  limit of a suitable sequential operation with a finite $t_L-t_0$.
On the other hand,  no
waiting time $t_l-t_{l-1}=0$ implies that the process $\lambda_{l-1} \to \lambda_l \to \lambda_{l+1}$ becomes a single quench process, and
 the amount of work in this process is given by $(\lambda_{l-1}-\lambda_{l+1})\bar u(\lambda_{l-1})$.

Here, we comment on generalizations of our arguments.
In the argument of  classical dynamics,  the condition on the confining potential can be relaxed.
The following conditions on $u(x)$
$$
\int_0 ^c \frac{\bar u(\lambda)-u(x)}{(E(\lambda)-\lambda u(x))^{\frac{3}{2}} }dx \geq 0, \ \  u(c) \leq u(c'),
$$
are sufficient for the validity of the inequality (\ref{2ndfinal}) to guarantee  the concavity of $E(\lambda)$.  
Confining potential $u(x)$ in wider class satisfies the above condition, and
amount of work in quasi-static operation $\lambda_0 \to\lambda_1$ is larger than that in the quench operation.
Although we have assumed  $u(-x) = u(x)$ for simplicity, the argument can be easily extend to general functions.

Finally, we discuss Maxwell's demon in this system.  The maximal work $W_{\rm max}(\lambda_0 \to \lambda_1)$ 
done by the particle in the quench operation $\lambda_0 \to \lambda_1$ is given by
\begin{equation*}
W_{\rm max}(\lambda_0 \to \lambda_1)=
\left\{
\begin{array}{ll}
(\lambda_0 -\lambda_1)u(\alpha(\lambda_0))
& (\lambda_0>\lambda_1)\\
0 &(\lambda_0\leq \lambda_1).
\end{array}
\right.
\label{Maxmax}
\end{equation*}
This is larger than the work 
$$
 W_{\rm qs}(\lambda_0\to\lambda_1)= E(\lambda_0) -E(\lambda_1),
$$ 
 in quasi-static operation $\lambda_0 \to \lambda_1$ as the maximal work in thermodynamics, since $E(\lambda)$ is convex and 
 $$
E'(\lambda_0) = \bar u(\lambda_0) \leq u(\alpha(\lambda_0)).
 $$
For $\lambda_0 > \lambda_1$, $W_{\rm max}(\lambda_0 \to \lambda_1)=(\lambda_0 -\lambda_1)u(\alpha(\lambda_0))$  is extracted by 
a quench expansion $\lambda_0 \to \lambda_1$
 when the particle is at $x=\pm \alpha(\lambda_0)$,  and for $\lambda_0 \leq \lambda_1$, no work $W_{\rm max}(\lambda_0 \to \lambda_1)=0$  is extracted by the quench 
 compression at the origin $x=0$.
 Note that the difference $$\Delta W:=W_{\rm max}(\lambda_0 \to \lambda_1)
-W_{\rm qs}(\lambda_0\to\lambda_1), 
$$ between the maximal  quench 
work and the quasi-static work 
is
 \begin{equation*}
\Delta W
=
\left\{
\begin{array}{ll}
\lambda_0  u(\alpha(\lambda_0)) - \lambda_1 u(\alpha(\lambda_1))
& (\lambda_0>\lambda_1)\\
\lambda_0  u(\alpha(\lambda_0)) - \lambda_1 u(\alpha(\lambda_1))
 &(\lambda_0\leq \lambda_1).
\end{array}
\right.
\label{Maxmax}
\end{equation*}
If Maxwell's  demon  can control each potential confining the corresponding particle,  the total amount of work 
exceeds the work in the quasi-static process.

Acknowledgments     

We are grateful to R. M. Woloshyn for a careful reading of the manuscript.
It is pleasure to thank T. Sako for helpful discussions.


\begin{thebibliography}{99}
\bibitem{PW}W. Pusz and S. L. Woronowicz, Comm. Math. Phys. {\bf 58}, 273 (1978).
\bibitem{L} A. Lenard, J. Stat. Phys. {\bf 19}, 575 (1978).
\bibitem{J} C. Jarzynski, Phys. Rev. Lett. {\bf 78}, 2690 (1997).
\bibitem{VR} L. Vidmar and M. Rigol, J. Stat. Mech. 064007 (2016).
\bibitem{DCHFGV}R. Dorner, S. R. Clark, L. Heaney, R. Fazio, J. Goold, and V. Vedral, Phys. Rev. Lett. {\bf 110}, 230601 (2013).
\bibitem{GRE} R. Gallego, A. Riera, and J. Eisert, New J. Phys. {\bf 16},125009 (2014).
\bibitem{PRGWE}  M. Perarnau-Llobet, A. Riera, R. Gallego, H. Wilming,and J. Eisert, New J. Phys. {\bf 18}, 123035 (2016).
\bibitem{MR} R. Modak and M. Rigol, Phys. Rev. E {\bf 95}, 062145 (2017).
\bibitem{LLMPP} T. P. Le, J. Levinsen, K. Modi, M. M. Parish, and F. A.Pollock, Phys. Rev. A {\bf 97}, 022106 (2018).
\bibitem{JSDMCG} F. Jin, R. Steinigeweg, H. De Raedt, K. Michielsen, M. Campisi, and J. Gemmer, Phys. Rev. E {\bf 94}, 012125 (2016).
\bibitem{SKCSG} D. Schmidtke, L. Knipschild, M. Campisi, R. Steinigeweg, and J. Gemmer, Phys. Rev. E {\bf 98}, 012123 (2018).
\bibitem{DPZ} S. Deffner, J. P. Paz, and W. H. Zurek, Phys. Rev. E {\bf 94} 010103(R) (2016).  
\bibitem{KIS} K. Kaneko, E. Iyoda and T. Sagawa, Phys. Rev. E {\bf 99}, 032128 (2019).
\bibitem{IA} C. Itoi and M. Amano, arXiv. 1911.01693.
\bibitem{LL}L. D. Landau and E. M. Lifshitz, Mechanics, 
Course of Theoretical Physics Volume 1 (Elsevier Butterworth-Heinemann, Oxford, UK, 1976).
\bibitem{H} H. Hellmann, Einf\"uhrung in die Quanlenchemie (Deuticke, Leipzig, Germany, 1937)
\bibitem{F}R. P. Feynman, Phys. Rev. {\bf 56}, 340 (1939).
\end{thebibliography}
\end{document}